# DNA driven self-assembly of micron-sized rods using DNA-grafted bacteriophage fd virions

R. R. Unwin†[a], R. A. Cabanas†[b], T. Yanagishima†[a], T. R. Blower[c], H. Takahashi[d], G. P. C. Salmond[c], J. M. Edwardson[d], S. Fraden[b], E. Eiser*[a,e]

[a] Cavendish Laboratory, University of Cambridge, JJ Thomson Ave, Cambridge, CB3 0HE, U.K.

[b] Martin Fisher School of Physics, Brandeis University, Abelson-Bass-Yalem 107, MS 057, 415 South Street, Waltham, MA 02453, U.S.A.

[c] Department of Biochemistry, University of Cambridge, Tennis Court Rd, Cambridge, CB2 1QW, U.K.

[d] Department of Pharmacology, University of Cambridge, Tennis Court Rd, Cambridge, CB2 1PD, U.K.

[e] BP Institute, Bullard Laboratories, Madingley Rd, Cambridge, CB3 0EZ, U.K.

† Equal contribution

* Corresponding author

We have functionalized the sides of fd bacteriophage virions with oligonucleotides to induce DNA hybridization driven self-assembly of high aspect ratio filamentous particles. Potential impacts of this new structure range from an entirely new building block in DNA origami structures, inclusion of virions in DNA nanostructures and nanomachines, to a new means of adding thermotropic control to lyotropic liquid crystal systems. A protocol for producing the virions in bulk is reviewed. Thiolated oligonucleotides are attached to the viral capsid using a heterobifunctional chemical linker. A commonly used system is utilized, where a sticky, single-stranded DNA strand is connected to an inert double-stranded spacer to increase inter-particle connectivity. Solutions of fd virions carrying complementary strands are mixed, annealed, and their aggregation is studied using dynamic light scattering (DLS), fluorescence microscopy, and atomic force microscopy (AFM). Aggregation is clearly observed on cooling, with some degree of local order, and is reversible when temperature is cycled through the DNA hybridization transition.

## 1. Introduction

Rod-like viruses have been widely used to study liquid crystal phenomena at the mesoscale [1], and have found widespread use due to their well-tuned monodispersity, mesoscopic length and ease of handling. Tobacco Mosaic Virus (TMV), further to its impact on virology [2], was one of the first rod-shaped viruses to be investigated in depth for its biophysical behaviour, inspiring the original works of Onsager on the liquid crystalline isotropic-to-nematic phase transition [3], with studies extending to the present day [4-6]. The more recent use of bacteriophages, produced by infected bacteria, provides a much simpler means to mass produce filamentous particles using standard culture facilities. In particular, M13 and fd phages have been extensively used to study liquid crystalline phase behaviour [7-12].

Another rapidly developing field is the DNA-driven self-assembly of colloids. Because of the specificity of Watson-Crick base pairing, an unprecedented level of control can be achieved in tuning inter-colloidal interactions by functionalizing the surface of these colloids with short, artificial oligonucleotides. In addition to the observation of amorphous aggregation [13-15], macroscopic crystallization has been achieved for nanoparticles with a vast range of morphologies [16-18]. Applications range from the addition of nanoparticles to DNA origami structures [19] to the development of novel biosensors [20]. A review of the physical properties and applications of these structures can be found in the literature [21, 22].

Here, we present a combination of these two innovative approaches to produce DNA driven self-assembly of mesoscopic filamentous particles. Work on the assembly of rod-like particles has been carried out previously [23,24] using gold nanorods, where crystallization was achieved through careful annealing; however, the aspect ratio and size of the particles was limited to the order of 1 to 5 due to the limited ability to synthesize high aspect ratio rod-like colloids [25]. The fd virions we use have a length of about 880 nm and a diameter of ~ 7 nm, hence their effective aspect ratio can be in excess of 100 when the negative charges of the coat proteins are sufficiently screened [26]. More importantly, filamentous viruses are highly monodisperse. Because of their large aspect ratio they are not completely rigid, but can be viewed as semi-flexible polymers, which can drastically change the rheological behaviour of aqueous solutions, with particular relevance to biological systems [27].

The potential applications of this work are many-fold. Firstly, there is a materials science interest, where these assembling structures can be incorporated into existing DNA origami architectures as a rigid, semi-flexible component, in the same manner as existing incorporations [28], or as a functional addition by application of phage-display methodologies [29]. Their size and flexibility also gives them a novel hydrodynamic behaviour in flow [12]. Secondly, viral capsids by themselves have been applied to the construction of nanoarchitectures [30]. The subsequent addition of DNA obviates the need to functionalize the viral capsid through genetic modification [31], broadening the complexity of the structures formed and significantly simplifying any protocol. Thirdly, in addition to the liquid crystalline properties of fd solutions [7-10], the addition of DNA to the viral surface enables a new type of temperature dependent interaction, introducing a whole new dimension of control to thermotropic liquid crystal systems, in that the melting temperature of the single-stranded (ss)DNA forms a sharp, finite transition point for a hybridization driven liquid crystal phase.

This article is primarily concerned with presentation of a protocol by which these structures can be produced, and visualization of some self-assembled structures formed through DNA hybridization. Specifically, we study a system in which the attached DNA consists of a 10 nm long double-stranded section (dsDNA), with a sticky, hybridizing ssDNA overhang. This DNA construct has been proposed as a more efficient functionalization tool for colloidal assembly compared with a single-stranded section [18, 21, 32], because once complementary ssDNAs have hybridized, the resulting short dsDNA behaves like a rigid rod, reducing the configurational entropy, and leading to a reduced binding strength between colloids.

## 2. Materials and Methods

### Mass production and purification of fd phage

In order to perform meaningful physical experiments to test the system's material properties, it is first necessary to review a method for fd production with high yield, high concentration, and low levels of impurities. We used a bacterial host E. coli ER2738 (New England Biolabs), a strain carrying the F' factor (conferring sensitivity to the male-specific fd bacteriophage) and expressing resistance to the antibiotic tetracycline (Tet), enabling positive selection and maintenance of the F' factor. A schematic of the procedure is given in Fig. 1A.

We began with an overnight starter culture taken from a single colony of ER2738 and incubated it in 10 ml of sterile 2xYT (yeast extract tryptone) growth medium at 37°C. For all experiments 2xYT broth was used throughout, unless stated otherwise. To ensure that we always used the same bacterial culture density, the resulting overnight E. coli culture was diluted into fresh growth medium to obtain an optical density (OD) of 0.5 at 600 nm, assayed using a UV-Vis spectrophotometer (see below). Subsequently, 3 ml of this diluted sample was inoculated with 10 µl fd phage stock at approximately 1 mg/ml, corresponding to $3.7 \times 10^{13}$ pfu /ml. After standing for 30 minutes to allow infection, the suspension was incubated again at 37°C for a further 30 minutes. The infected culture was transferred to a 250 ml conical flask with 30 ml of medium and incubated on an orbital shaker at 250 rpm and 37°C for 2 hours. Finally, we transferred the medium-sized culture to a large 2 L flask with 500 ml of medium, and incubated overnight on an orbital shaker (250 rpm, 37°C).

Purification was achieved as follows: E. coli cells and debris were removed by centrifuging the cultures twice at 8,300 g for 15 minutes, harvesting the supernatant each time. A precipitant solution was added (146.1 g/l sodium chloride + 200 g/l PEG 8,000) in a ratio of 3 parts precipitant to 10 parts supernatant. After refrigeration for at least an hour, the supernatant was centrifuged as before, and the clear supernatant was carefully removed, leaving the precipitated viral mass in the pellet. We resuspended the viral pellet in 10 ml of sterile phosphate buffered saline (PBS) solution. The concentrated viral solution was produced after further centrifugation to remove any residual debris (10 minutes at 4,000 g). Ultra-centrifuge tubes were prepared containing a caesium chloride (CsCl) gradient with an average specific gravity of 1.4 and the viral precipitate was layered on to the top CsCl solution. There was typically 1 part supernatant to 3 parts CsCl solution. This mixture was then spun at 180,000 g in an ultracentrifuge at room temperature for approximately 48 hours. After density equilibration, the tubes were removed and the phage band was carefully collected with a fine-tipped pipette. The phage sample was then dialysed against sterile PBS and stored at 4°C until further use. An AFM image of the wild type fd and an optical microscopy image of 27 mg/ml fd in PBS under crossed polarizers are shown in Fig. 1B and 1C, for reference.

### Oligonucleotide functionalization

The fd viral capsid is made up of approximately 2700 copies of the major coat protein pVIII, each with a hydrophilic primary amine group pointing outwards [33]. These coat proteins form a hollow cylindrical tube containing the single-stranded viral DNA. We used a heterobifunctional linker, Sulfo-SMCC (Sulfosuccinimidyl-4-   (N-maleimidomethyl) cyclohexane-1 carboxylate) (Sigma Aldrich), which contains both a succinimidyl ester and a maleimide group, to covalently bind our thiolated DNA to the endpoints of pVIII. Sulfo-SMCC readily attaches to primary amine groups, while the latter can link to thiol groups. We detail here material quantities required to label 0.5 mg of fd with DNA. The amount of thiolated oligo required for this amount can be significant. We thus scaled up the protocol of Hermanson [34] to thiolate unlabelled oligonucleotides (Integrated DNA Technologies). Details of the reaction pathway are given in [34].

A tube of (nominally) 1 µmole of oligo was reconstituted in 100 µl of saline HEPES (150 mM sodium chloride in 10 mM 4-(2-hydroxyethyl)-1-piperazineethane sulfonic acid). Half of this was used, and the other half stored at -20°C for future use. 8.3 mg of EDC (1-ethyl-3-(3-dimethylaminopropyl)carbodiimide) and 33 µl of

0.25 M cystamine dihydrochloride in pH 6.0 0.1 M imidazole was added to 50 μl of DNA solution, and vortexed thoroughly. After another 133 μl of imidazole was added, the mixture was allowed to react for 30 minutes. Unreacted compounds were removed by double desalting using two Zeba Desalting Columns MWCO 7k.

Three oligo sequences were used in this work: dA, dAP, and InertC; their sequences are given in Table 1. dA and dAP were thiolated at the 5' end, and subsequently annealed in an equimolar ratio with InertC at 80°C in a heating block for 1 hour. The sample was then allowed to cool slowly overnight, giving rise to a thiol connection to the capsid, a group of 5 thymines for flexibility, followed by a 10 nm long inert double stranded spacer of 30 base pairs, another 5×T group, and a 10 base pair overhang with a free 3' end, for which dA and dAP carry complementary sequences, in that order.

After the DNA structure was formed, the linkage procedure (outlined in Fig. 2 A) was as follows: 2 mg of Sulfo-SMCC was placed in a 1.5 ml tube was dissolved in 1 ml of 10 mM phosphate buffer. It should be noted that we used phosphate buffer in this step, because Sulfo-SMCC is only poorly soluble when added directly into high salinity buffers such as PBS, pH = 7. 10 μl of a 50 mg/ml stock of highly concentrated fd solution was added, to produce a mixture of 0.5 mg/ml of fd virions and a 50 times molar excess of linker to the number of binding sites to be reacted. This mixture was allowed to react for one hour, before removing excess linker by a double desalting procedure using two Zeba Desalting Columns MWCO 40k (Thermo Scientific).

The above thiolated (and hybridized) DNA structure carries a disulfide bridge, which requires reduction before the thiol group becomes available. For this reduction, 133 μl of 1M TCEP-HCl in ddH$_2$O (Tris(2-carboxyethyl)phosphine hydrochloride, Thermo Scientific) was added, and allowed to react for 15 minutes; then the solution was mixed with the maleimide-carrying virus. This approach made use of one of the key benefits of TCEP - its relatively low competition with any thiol-maleimide coupling. The fd virions were allowed to react with the thiol-DNA for 2 hours, before overnight dialysis against 1L batches of saline HEPES, replacing the volume every few hours.

### Agarose gel electrophoresis

As a preliminary evaluation of the binding of DNA to the fd, agarose gel electrophoresis was carried out with uncoated and coated virus in parallel channels. A 10 x 25 cm 1.5% (w/w) agarose gel in TAE (Tris-acetate ethylenediaminetetraacetic acid) buffer was produced by standard means, and loaded with 10 μl samples, consisting of a 1:3 ratio of bromophenol blue loading dye and ~ 0.1 mg/ml of virus. Electrophoresis was carried out at 50 V for 4 hours. Subsequently, the protein capsid of the virions was dissolved using 250 mM sodium hydroxide, enabling the visualization of the viral DNA using ethidium bromide. Care was taken to ensure that any base or dye was thoroughly removed by rinsing with TAE buffer before the next step. Finally, the gel was observed in a Gel Doc-It (UVP LLC) imaging box.

### Fluorescence microscopy

To demonstrate the binding of DNA to the virions, we used the same protocol to attach a ssDNA oligomer (IDT technologies) to the surface of the virus. The oligomer, labelled sG, was composed of a thiol group, followed by an 14 thymine base spacer, and ending with a 10 base sticky end; see Table 1 for the sequence. In addition to this, we used another ssDNA (IDT technologies) tagged with a fluorophore. This strand, labelled F, possessed a Cy3 fluorophore at the 5' end, followed by a 10 base single strand DNA known as a 'sticky end', and complementary to the sG strand (see Table 1 for details). The Cy3 fluorophore has an excitation wavelength of 550 nm and an emission wavelength of 570 nm. This is nearly identical to TRITC (tetramethyl rhodamine isothiocyanate), but brighter, more photostable and with less background fluorescence. We used a TRITC filter set (Chroma) to visualize the fluorescence on an inverted microscope (NIKON Eclipse Ti), equipped with a high numerical aperture oil objective (NIKON 100x Plan-Fluor NA 1.3) and a Mercury Halide lamp (X-Cite series 120Q). Images and a movie were recorded using a high sensitivity interline CCD camera (Andor Clara, ICX285 sensor from Sony; see supporting movie). Exposure time was kept at the maximum possible without blurring the image (120 ms), due to the movement of fd virions in solution.

The DNA-fluorophore (F) and fd-DNA (fd-sG) hybrids were mixed in 100 mM PBS solution at a ratio 400:1, and 0.01 mg/ml virus concentration. The mixture was placed in a chamber made with a 25 x 75 mm microscope slide and 18 x 18 mm coverglass (Corning), with a piece of Parafilm (Pechiney Plastic Packing) with a hole punctured in the middle acting as a space holder and cavity between the two glasses.

### UV-Vis spectrophotometry

A Nanodrop 2000 UV-Vis spectrophotometer (Thermo Scientific) was used to measure virus concentrations and the concentrations of any DNA samples, when required, using extinction coefficients of 3.84 $A_{269nm}$ (mg/ml)-1 cm-1 for fd, and 0.020 $A_{260nm}$ (μg/ml)$^{-1}$ cm$^{-1}$ for DNA.

The absorption spectra for fd and DNA can be expressed as a function of wavelength λ, using the Beer-Lambert law,

$$A_{fd}(\lambda) = LC_{fd}\epsilon_{fd}(\lambda), A_{DNA}(\lambda) = LC_{DNA}\epsilon_{DNA}(\lambda),$$

where Afd and ADNA are the absorption spectra for fd and DNA solutions respectively, L is the path length of absorption, Cfd and CDNA¬ are the concentrations of fd and DNA respectively, and ε represents the respective wavelength dependent absorptions for each material per unit path length, and unit concentration. By measuring separate solutions of fd and DNA at known concentrations, it was possible to obtain unique values for ϵ for each material.

It is hypothesized that when DNA is attached to the fd capsid, the DNA and fd absorption profiles are additive. Neglecting the small amount of absorption attributed to the newly formed SMCC linkage, we quantified the amount of DNA bound per fd virion by treating the hybrid structure as a superposition of the fd and DNA wavelength dependent absorption curves from 220 nm–350 nm,

$$A_{fd+DNA}(\lambda) = L[C_{fd}\epsilon_{fd}(\lambda) + C_{DNA}\epsilon_{DNA}(\lambda)].$$
(1)

In equation (1) there are two unknowns; the fd and DNA concentrations. Because the absorption spectrum is measured in increments of 1 nm, there are over 100 data points and thus the concentrations are over-determined. The amount of DNA bound per virus is obtained in a least-squares fit of the measured absorption of the fd/DNA particle to eqn. (1).

**Atomic force microscopy (AFM)**

Since the fd virion is only about 8 nm wide, it is not possible to visualize its diameter using optical means. We therefore used AFM to visualize wild type fd, DNA coated fd, and any DNA-driven aggregates of fd-virions grafted with complementary ssDNA overhangs. The resolution was such that even individual strands of dsDNA could be visualized [35]; however, when attached to a significantly larger structure such as the virion, the dsDNA could not be visualized directly.

Samples were diluted to approximately 5 μg/ml in 20 mM HEPES-NaOH buffer, pH 7.6, containing calcium chloride at 1 mM to induce adhesion to the mica substrate. It should be noted that this procedure itself may cause weak non-specific fd aggregation, and so a control image is shown for comparison (Fig. 4A). The surface of a 5 mm diameter mica disc on a 15 mm steel SPM specimen holder (Agar Scientific) was freshly cleaved using adhesive tape (Aron Alpha type 102, Agar Scientific). The sample (50 μl) was pipetted gently onto the surface and allowed to adhere to the mica for three minutes. The sample was gently washed with Biotechnology Performance Certified water (Sigma) and was then dried rapidly using nitrogen gas before imaging.

Imaging was carried out at room temperature in air using a Bruker Multimode atomic force microscope controlled by a Nanoscope IIIa controller with an E-scanner using tapping mode. The silicon nitride probes (OTESPA; Bruker AFM Probes) typically had a drive frequency of ~300 kHz and a specified spring constant of 40 N/m, and were tuned to 10-20 Hz below the resonance frequency peak. The applied imaging force was kept as low as possible ($A_S/A_0$~0.85). Images were captured with 512x512 scan lines per image at a scan rate of 2 Hz.

**Dynamic light scattering**

As a parallel means of measuring aggregation, a Zetasizer Nano ZS (Malvern Instruments) was used to observe the hydrodynamic size of fd-virus solutions, as a function of temperature. Samples consisting of 50 μl of 0.5 mg/ml aliquots of dA-fd and dAP-fd were preheated to 50°C, and added to 900 μl of preheated PBS, producing a sample with a total of 0.05 mg/ml fd. This sample was introduced into a semi-micro cuvette rinsed with buffer, and the temperature was decreased in 0.5°C steps, allowing 5 minutes of equilibration at each temperature.

In a mixture of fd coated with dA and dAP starting at high temperature, the onset of aggregation was expected as the sample was slowly cooled through the melting temperature Tm of the sticky ends. In 150 mM salt (PBS salinity), our 10 base pair sticky ends have a Tm ≈ 36°C, calculated using the pairwise interaction formula of SantaLucia [36]. We therefore studied a range of temperatures centred on this value, 20–50°C.

We expected a highly polydisperse sample, with free fd-virions mixed with significantly larger aggregates. A non-negatively constrained least-squares (NNLS) method was used, which explicitly gives the proportion of the scattered signal attributed to each different aggregate size.

## 3. Results

**UV-Vis absorption of fd-DNA hybrid structures**

We employed UV-Vis absorption to test the efficiency of our protocol to covalently bind DNA to fd-virions. We measured the absorption spectrum of 0.5 mg/ml fd grafted with inertC-hybridized dAP oligos. The absorption spectrum is shown in Fig. 3A, along with spectra for unlabelled fd and pure DNA solutions. The absorption spectrum of the fd/DNA complex shows attributes that are unique to the pure fd and pure DNA spectra. The absorption of the virus-DNA complex and the result of a least squares fit are compared in Fig. 3A: excellent agreement is seen between them. Fitting reveals an fd concentration of 0.283 mg/ml (17.7 nM) and a DNA concentration of 0.0262 mg/ml (1050 nM). The fd concentration is 40% lower than the initial concentration, which may indicate losses in dialysis, sample transfer and desalting. The concentration of DNA bound to fd virus corresponds to roughly 70 DNA strands grafted to each fd virion.

**Agarose gel electrophoresis of fd-DNA hybrid structures**

We also ran the coated virions on an agarose gel using standard electrophoretic techniques, and compared them with uncoated virions. Fig. 3B shows migration distances; the coated virions showed reduced mobility compared with that of the uncoated versions, as expected from the increase in their hydrodynamic radius and the drag from the gel matrix due to the presence of DNA protrusions from their surfaces.

### AFM imaging – control dA-fd vs. dA-fd + dAP-fd annealed sample

Having established the success of the grafting procedure, a solution containing a 1:1 ratio of dA-fd to dAP-fd at approximately 5 µg/ml was heated to 50°C for 1 hour, well above the melting temperature of the sticky ends, Tm ≈ 36°C. This solution was then allowed to cool slowly overnight, and the sample was then imaged by AFM. AFM images of a dA-fd sample that was used as control and of the annealed fd+dA/fd+dAP mixture that was allowed to hybridize are shown in Fig. 4A and B, respectively. Both samples showed small clusters of fd virions, whose adhesion is likely due to calcium ions present in the buffer solution used to deposit the virions onto the mica substrate. However, the annealed sample containing the virions with complementary DNA strands (Fig. 4B) revealed large, extended structures that were strikingly different from those in the control sample (Fig. 4A). A wide range of geometries could be observed, including loops (Fig. 4C), dendritic (Fig. 4D), and extended dendritic structures (Fig. 4E).

### Fluorescence imaging of DNA functionalized fd

Fluorescence imaging was performed to test whether DNA is indeed attached to the virions. For this we used fd-virions coated with the sG oligo, which carries the same sticky overhang as the dA strand. However, this sticky strand is attached to the virion via a shorter single stranded spacer of 14 thymines. At temperatures above the melting point of the DNA strand F, the unhybridized fluorescently-tagged oligomer diffuses through the sample much more quickly than the exposure time of the camera, producing a uniform fluorescence background caused by the cumulative effect of all diffusing oligomers.

As the temperature decreases, the tagged oligomers start hybridizing to the complementary strands of DNA attached to the viral capsids. For temperatures sufficiently lower than the melting temperature of the DNA (~36°C), sufficient fluorophore-labelled DNA has hybridized onto the virions for them to shine more brightly than the background. Furthermore, the diffusion of the virus in solution is much slower than that of the oligomer alone, enabling one to take clear movies of the diffusion of the virus when covered by fluorophore tagged oligomers (see supporting video). This set-up facilitates the detection of individual virions, since the concentration of the virus is such that a few virions will be in the microscope's field of view at once, but not so many that they cannot be distinguished from the background.

As seen in Fig. 5, the fluorescent oligomers started to detach from the virus as the temperature increased, and, above the melting temperature, we recovered the initial uniform fluorescent background. This process can be cycled several times, obtaining the same result as in the first cycle.

### DLS of dA-fd + dAP-fd aggregation

The results of DLS measurements of fd+dA/fd+dAP mixtures at different temperatures, using a dilute solution of bare fd virions as reference, are shown in Fig. 6. By choosing sufficiently dilute solutions in all cases, we ensured that in the non-hybridized state, the effective excluded volume of the individual virions did not influence their free diffusion. The bare virions display two characteristic relaxation times, corresponding to related length scales typical for semi-flexible polymers with large aspect ratio [37,38]. At temperatures above Tm, the DNA-functionalized virions also display two peaks, albeit shifted to slightly higher relaxation times due to the increased molecular weight. The bimodal appearance of the distribution of the characteristic length scales is shown in Fig. 6. The time-correlation function of the scattering intensities measured in polarized DLS (as used here) contains two terms: (1) one that depends on the translational diffusion coefficient of the rods and the square of the scattering vector q, and (2) a second term that depends on the rotational diffusion and internal modes related to the thermal vibrations of the rod ends. The latter is usually independent of q. When we cooled the sample to about 39°C, an additional peak appeared at a significantly larger size, in excess of 880 nm, the length of the fd virion accounted for by the second peak. This new peak persisted as the temperature was further lowered, and the peaks at smaller hydrodynamic radii diminished in intensity suggesting that single virions were incorporated into the clusters. It is interesting to note that the two peaks of the free virions were still present in the larger aggregates but shifted to larger sizes. This is not surprising as the aggregates formed have few binding contacts between each other (see also the AFM images in Fig. 4), while many rod ends still remain 'free', most likely contributing to the signal at small sizes. In addition, a small fraction of free fd virions will probably remain in solution, even after several hours of aggregation. At lower temperatures, the signal became significantly noisier, owing to heterogeneity and high polydispersity in the aggregating sample. Finally, it should be noted that larger aggregates will scatter more towards smaller scattering angles, while all experiments were performed at a single scattering angle of 90°.

## 4. Discussion

Given 2700 available sites on each virion, roughly 70 bound DNA strands seems to be a very small number. Indeed, much higher grafting densities can be achieved by adjusting the salinity of the buffer as well as the initial DNA concentration in the grafting process. In fact, for the fluorescence experiments, where we used ssDNA spacers, a grafting density of ~ 400 DNA strands per virion was used (Fig. 5). However, for the sake of monitoring the aggregation process in time, we chose to work with low grafting densities. In order to understand the slow aggregation process observed in our experiments, it is important to recognize that the binding or hybridization transition of binary mixtures of short complementary ssDNA solutions is not a sharp one. Using the absorption spectrum of DNA at 260 nm as a function of temperature, strong absorption is seen above the melt temperature where all DNA pairs are unbound [21]. As the temperature is lowered, one observes a steady decrease of the absorbed intensity until a plateau is reached at which all DNA strands form duplexes. The inflection point of this s-shaped curve denotes the melt temperature. At this point half of the strands are bound and the other half unbound. The width of the transition can be 20-30°C. Furthermore, the hybridization process is slow, taking place over the course of hours. For the same DNA sequences and concentration, but end-grafted to spherical colloids, a dramatic narrowing of the transition occurs, which can become as small as 1°C. Moreover, the melt temperature increases logarithmically as a function of possible hydrogen bonds in the colloid's contact area. These observations are due to a cooperative effect introduced by the locally increased DNA concentration on the colloidal surface, as discussed by Geerts and Eiser [21], and Lukatsky and Frenkel [39]. Such a narrow melt transition also explains the rapid aggregation (within seconds) that was reported for the case of densely grafted spherical colloids [14,15]. Our system displays an intermediate behaviour with slightly increased melt temperature and aggregation dynamics.

To illustrate the distribution of DNA along the fd virions, we considered the area swept out by the rigid, inert dsDNA spacers attached via a short ~1.5 nm long and flexible T-spacer, which allows the double-stranded rod to pivot on the coat protein. The 36 bases of the dsDNA have a length of roughly 10 nm: hence, it can sweep out a hemisphere with a comparable radius, reducing the probability of subsequent DNA-functionalization. Assuming an effective radius of gyration of 12 nm associated with the entire length of the DNA strands used, only about 50 can fit along a single filament. The number will of course be slightly higher, as the DNA strands can bind not only along one dimension but all around the cylindrical virion. Higher grafting densities can be achieved, though in this case the salinity of the buffer solution must be increased, thereby reducing the Debye screening-length or Coulomb repulsion between the negative charges along the DNA backbone. The range of this repulsion is of the order of a few nanometers for the electrolyte (buffer) concentrations used.

Our AFM measurements confirmed that there is a DNA-driven aggregation of filamentous virions through the hybridization interaction. Particularly interesting is the range of geometries revealed by the AFM images (Fig. 4). There are confined, 'fibrillar' regions of order, where the fd virions clearly align; there are loops that form from complementary rods joining with minimal overlap, such that their semi-flexibility allows the opposite ends to also join; and there are dendritic structures, where fibrillar regions with poor overlap can split and accept other complementary strands. One notable absence is that of extended domains of nematic order. This is due in part to the fact that we worked at fd concentrations below the isotropic to nematic region. However, one might have expected some nematic regions to form because of the relatively high binding Gibbs free energy between individual active DNA pairs $\Delta G^0(dA/dAP) \approx -10.2$ kcal/M, which corresponds to about -17.2 $k_B T$ at room temperature ($k_B$ is Boltzmann's constant and the values correspond to our solvent conditions [36]). Similar values hold for the identical pair sG/F used in the fluorescence experiments discussed later. With about 70 DNA strands per fd virion, parallel aggregation would result in a very strong binding corresponding to a minimum in the Gibbs free energy. Even if not all complementary DNA strands on opposing fd-rods could be realized, since we cannot assume an equally spaced distribution of the DNA along the rods, strong binding between the rods takes place. A key to understanding the origin of the 'open' aggregates lies in the kinetics of the process. Similar to the aggregation of DNA functionalized spherical colloids [16], the main obstacles to attainment of equilibrium crystallization is dynamic arrest: once a pair of complementary DNA strands on two different particles bind, the probability that all 10 base pairs will subsequently unbind simultaneously is very low and decreases exponentially with each additional hybridized DNA duplex [39, 40]. Hence, rearrangement or 'annealing' to a lower free energy state is kinetically hindered. In other words, the time needed to reach the ordered equilibrium state vastly exceeds realistic experimental times.

DLS measurements support our findings that the aggregation process leads to a random, strongly disordered system of DNA-linked fd virions. Firstly, our results show clear evidence that the aggregation process being driven specifically by hybridization of the sticky DNA ends, as manifested in the appearance of longer and longer relaxation times, reflecting the strong binding between a few points along the rods (Fig. 6). Aggregation commences at about 40°C, which is somewhat above the calculated thermodynamic melting temperature (Tm ≈ 36°C). This is in accordance with our assumption that the melt transition has narrowed but still covers some 10°C [32], being between the limits of completely free DNA strands in solution and the sharp transition observed for densely coated spherical particles. Further, the DLS data show that the relaxation modes of the vibrating free rod-ends (around 10 nm) and the translational mode of the free rods (~ 880nm) are still present once aggregation sets in. This represents further evidence that an open network of semi-flexible rods forms, which is connected only via a few points, similar to a cross-linked actin filament gel.

Finally, we also need to comment on the fluorescence microscopy data presented in Fig. 5. These measurements were done as an independent control demonstrating the successful binding of DNA to the virions using the protocol detailed in the Materials section. Using higher salt concentrations during the attachment of thiolated DNA, a grafting density of roughly 400 sG oligos per virion was achieved. The higher grafting density was intended to enable better visualization of the hybridization between the sG-coated virions and the fluorescently labelled F oligos. For the fluorescence microscopy experiments the final salt concentration of the sG-virion/F mixture was adjusted to 50 mM NaCl. This reduced salt concentration shifted the melt temperature of the duplex to about 28 ° C. Starting recording fluorescence images at 32°C (Fig. 5), we saw few fluorescent specks, indicating that some F-oligos had hybridized to some virions. As the temperature was lowered, more and more fluorescent spots appeared, reflecting the increased number of F-oligos condensing onto the fd-virons, as expected. Note that unbound F-oligos only contribute to the overall averaged low intensity background as they diffuse much more quickly than the viruses. When the sample is heated back to well above Tm, the images become homogenously grey, reflecting the fact that all duplexes are molten. A video demonstration of this process is available as supplementary movie. The system was so robust that many hybridization-melt cycles could be repeated, always showing the same temperature behaviour.

## 5. Conclusion

We have developed a successful protocol for grafting oligonucleotides to the side-coat proteins of fd filamentous viral capsids, and, using AFM, we have visualized DNA-driven aggregates of these rod-like particles into a variety of geometries. We have detailed a protocol for the mass production of fd, the pre-thiolation of the oligo, and subsequent functionalization of the capsid. An entirely analogous temperature sensitivity to DNA coated spherical colloids was identified by DLS, with a significantly wider melting transition attributed to the range of rod-rod bonds involved in the interactions. Not only does this hybrid system have the potential to allow custom tuning of melting transitions, it can be incorporated into other DNA colloid systems, such as origami structures and biosensors.

## 6. Acknowledgements

TY acknowledges the support of the Ernest Oppenheimer Fund, the George and Lillian Schiff Foundation, and the Gonville and Caius College Travel Grant for financial support. RU acknowledges the Engineering and Physical Sciences Research Council (EPSRC) for financial support. Work in the GPCS Lab is funded by the BBSRC, UK. EE thanks the Winton Program for the Physics of Sustainability. SF and RAC acknowledge support from the US National Science Foundation MRSEC DMR-0820492 and SF acknowledges support from the Leverhulme Trust. HT held a Newton International Fellowship.

## Tables

| dA | HS-5'-TTT TTA AGG GAG AAA AGA GAG AGG GAA AGA GGG AAT TTT TTC TAA GCC AT-3' |
|---|---|
| dAP | HS-5'-TTT TTA AGG GAG AAA AGA GAG AGG GAA AGA GGG AAT TTT TAT GGC TTA GA-3' |
| inertC | 5'-TTC CCT CTT TCC CTC TCT CTT TTC TCC CTT-3' |
| sG | HS-5'-TTT TTT TTT TTT TTT CTA AGC CAT-3' |
| F | Cy3-5'-AT GGC T TA GA-3' |

**Table 1** List of DNA sequences used in this work. dA and dAP are annealed with inertC to give an inert dsDNA spacer terminated with a sticky ssDNA overhang of 10 bases. sG comprises a 14 thymine base spacer, followed by a 10 base pair sticky end, which hybridizes with the 10 base pair fluorophore tagged F strand.

## Figures

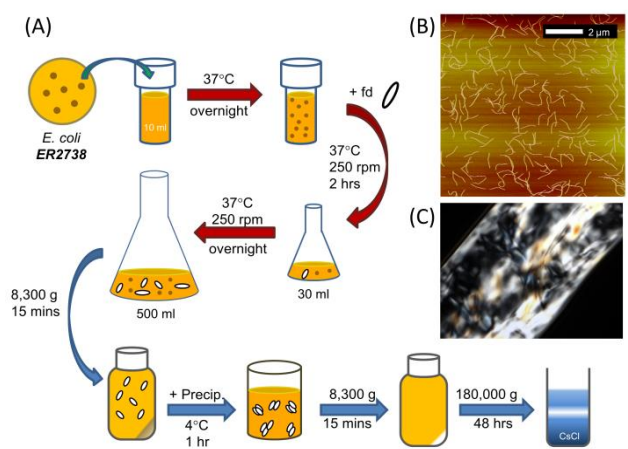

**Fig. 1** (A) Illustration of the virus production protocol. (B) Wild type fd (0.005 mg/ml) imaged on a mica surface, observed using AFM. (C) fd virus in PBS (25 mg/ml), observed under crossed polarizers inside a glass capillary. Significant birefringence is observed.

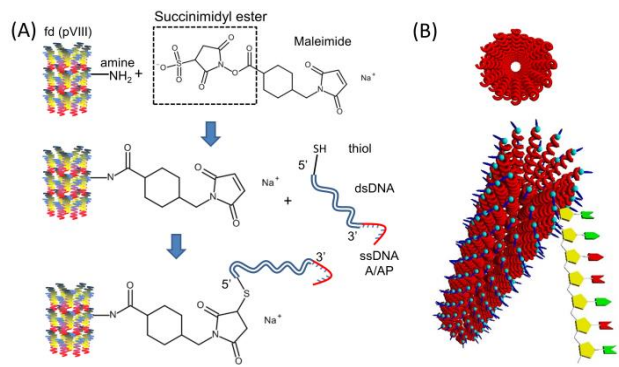

**Fig. 2** (A) Schematic of heterobifunctional linkage of oligos to the fd pVIII coat protein. (B) 3D representation of the top view of a wild type fd-virion, showing its hollow inner part, and a side view showing an individual ssDNA with arbitrary sequence linked to major coat protein pVIII.

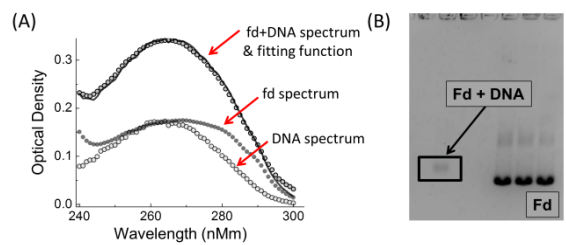

**Fig. 3** (A) UV-Vis spectra of wild type fd, DNA and fd+DNA hybrid structures taken at room temperature (here the sG/F duplex was used). A 0.5 mg/ml sample of wild type virus and a $\approx$ 0.5 mg/ml sample of fd+dA hybrid were compared. A least-square fitting of the absorbance $A_{fd+DNA}(\lambda)$ provides a grafting density of approximately 400 sG-oligos per virion. (B) Image of an agarose gel: the three dark bands on the right show the bare fd-solutions (running from top to bottom). The faint band on the left shows the DNA-functionalized fd virions, which have a considerably higher molecular weight.

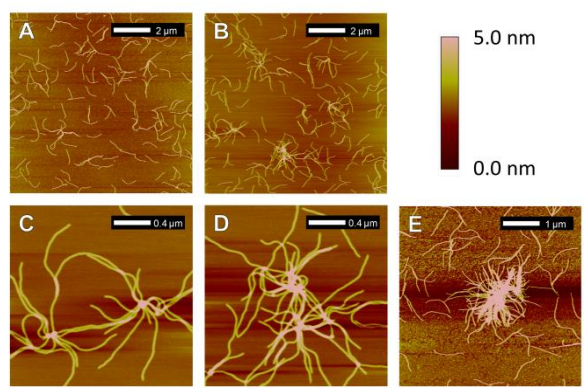

**Fig. 4** AFM images of fd+dA (A) and fd+dA/fd+dAP (B) annealed structures. The dA+dAP sample is clearly aggregated into larger, extended structures. Loops (C), dendritic formations (D) and radial structures (E) are also seen.

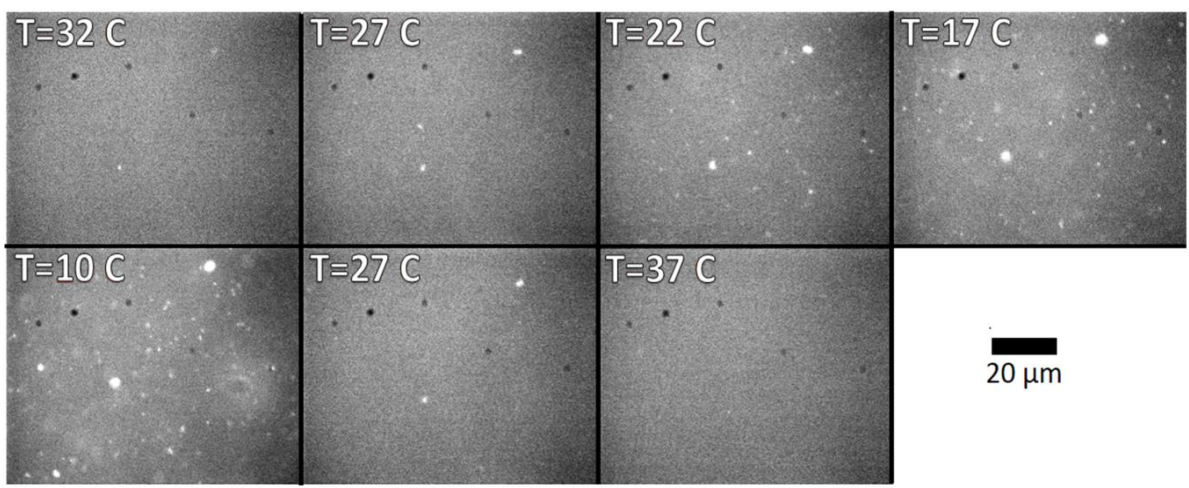

**Fig. 5** Fluorescence microscopy images taken for solutions containing sG-functionalized fd virions ($\approx$ 5 ng/ml, with 50 mM added NaCl) and the fluorescently labelled complementary F-strand. The images were taken at different temperatures starting above the melt temperature of the sG/F duplex, which was about 28°C in the solvent conditions used. The fluorescent specks are virions with hybridized DNA duplexes. Subsequently the sample was re-heated to 39°C. The images read from left to right and from top to bottom row.

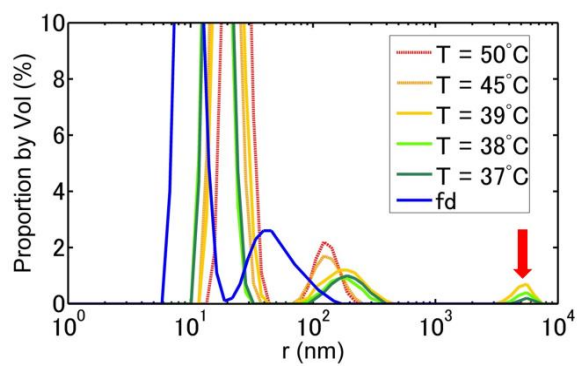

**Fig. 6** DLS aggregate size proportions (per volume) for pure dilute fd-solutions and for an equimolar mixture of dA and dAP-grafted virions at different temperatures; Tm ≈ 36°C. Both pure and DNA-functionalized fd virus above the melt temperature show two peaks. Upon cooling below about 40°C the mixtures show an additional peak (indicated by the red arrow) at much larger aggregate sizes that persists at lower temperature.